\documentclass[conference]{IEEEtran}
\IEEEoverridecommandlockouts
\usepackage{cite}
\usepackage{amsmath,amssymb,amsfonts}
\usepackage{algorithmic}
\usepackage{graphicx}
\usepackage{textcomp}
\usepackage{xcolor}
\usepackage{multirow}
\usepackage{float}
\usepackage{acronym}
\newacro{KV}{key-value}
\newacro{LLM}{large language model}
\newacro{PIM}{processing-in-memory}
\newacro{LoRA}{low-rank adaptation}
\newacro{PE}{processing element}
\newacro{IPCN}{inter-PE computational network}
\newacro{SRPG}{SRAM reprogramming and power gating}
\newacro{CT}{compute tile}
\newacro{DMAC}{dynamic data multiply-accumulate}
\newacro{SMAC}{static weight multiply-accumulate}
\newacro{RRAM-ACIM}{resistive RAM analog compute-in-memory}
\newacro{SRAM-DCIM}{static RAM digital compute-in-memory}
\newacro{NMC}{network main controller}
\usepackage{cleveref}
\crefname{figure}{fig.}{figs.}
\def\BibTeX{{\rm B\kern-.05em{\sc i\kern-.025em b}\kern-.08em
    T\kern-.1667em\lower.7ex\hbox{E}\kern-.125emX}}
\begin{document}

\title{PRIMAL: \underline{Pr}ocessing-\underline{I}n-\underline{M}emory based Low-Rank \underline{A}daptation for \underline{L}LM Inference Accelerator\\
}

\author{\IEEEauthorblockN{Yue Jiet Chong$^1$\textsuperscript{\dag}, Yimin Wang$^2$\textsuperscript{\dag}, Zhen Wu$^3$, Xuanyao Fong$^4$}
\IEEEauthorblockA{Department of Electrical and Computer Engineering, National University of Singapore, Singapore \\
Email: \{jason.yj.chong$^1$, kelvin.xy.fong$^4$\}@nus.edu.sg, \{yimin.wang$^2$, e0323083$^3$\}@u.nus.edu}
\thanks{This work is funded in part by the National University of Singapore through the Microelectronics Seed Grant (FY2024); and in part by the National Research Foundation (NRF), Singapore, under the Competitive Research Programme (Award NRF-CRP24-2020-0002 and NRF-CRP24-2020-0003).
}
\thanks{\textsuperscript{\dag} Both authors contributed equally to this work.}
}

\maketitle

\begin{abstract}
This paper presents PRIMAL, a \ac{PIM} based \ac{LLM} inference accelerator with \ac{LoRA}.
PRIMAL integrates heterogeneous \ac{PIM} \acp{PE}, interconnected by 2D-mesh \ac{IPCN}.
A novel \ac{SRPG} scheme enables pipelined \ac{LoRA} updates and sub-linear power scaling by overlapping reconfiguration with computation and gating idle resources.
PRIMAL employs optimized spatial mapping and dataflow orchestration to minimize communication overhead, and achieves $1.5\times$ throughput and $25\times$ energy efficiency over NVIDIA H100 with \ac{LoRA} rank 8 (Q,V) on Llama-13B. 
\end{abstract}


\section{Introduction}

Recent advancements in \acp{LLM} have significantly enhanced their ability to perform complex tasks, leading to widespread adoption across various domains such as customer service, software engineering, and content generation~\cite{llm_in_industry}.
Despite their general capabilities, models like GPT and LLaMA often fall short in meeting the specific requirements of downstream tasks, necessitating task-specific adaptation to improve performance and relevance~\cite{llm_pre_vs_post_adapt}.
Model adaptation enhances task alignment and offers a means to mitigate data privacy concerns by enabling on-premise or controlled fine-tuning.
However, full model fine-tuning is computationally intensive and resource-demanding, making it impractical for many applications.
To address this limitation, \ac{LoRA}~\cite{lora_origin} has emerged as an efficient alternative. 

\begin{figure}[t]
    \centering
    \includegraphics[width=0.75\linewidth]{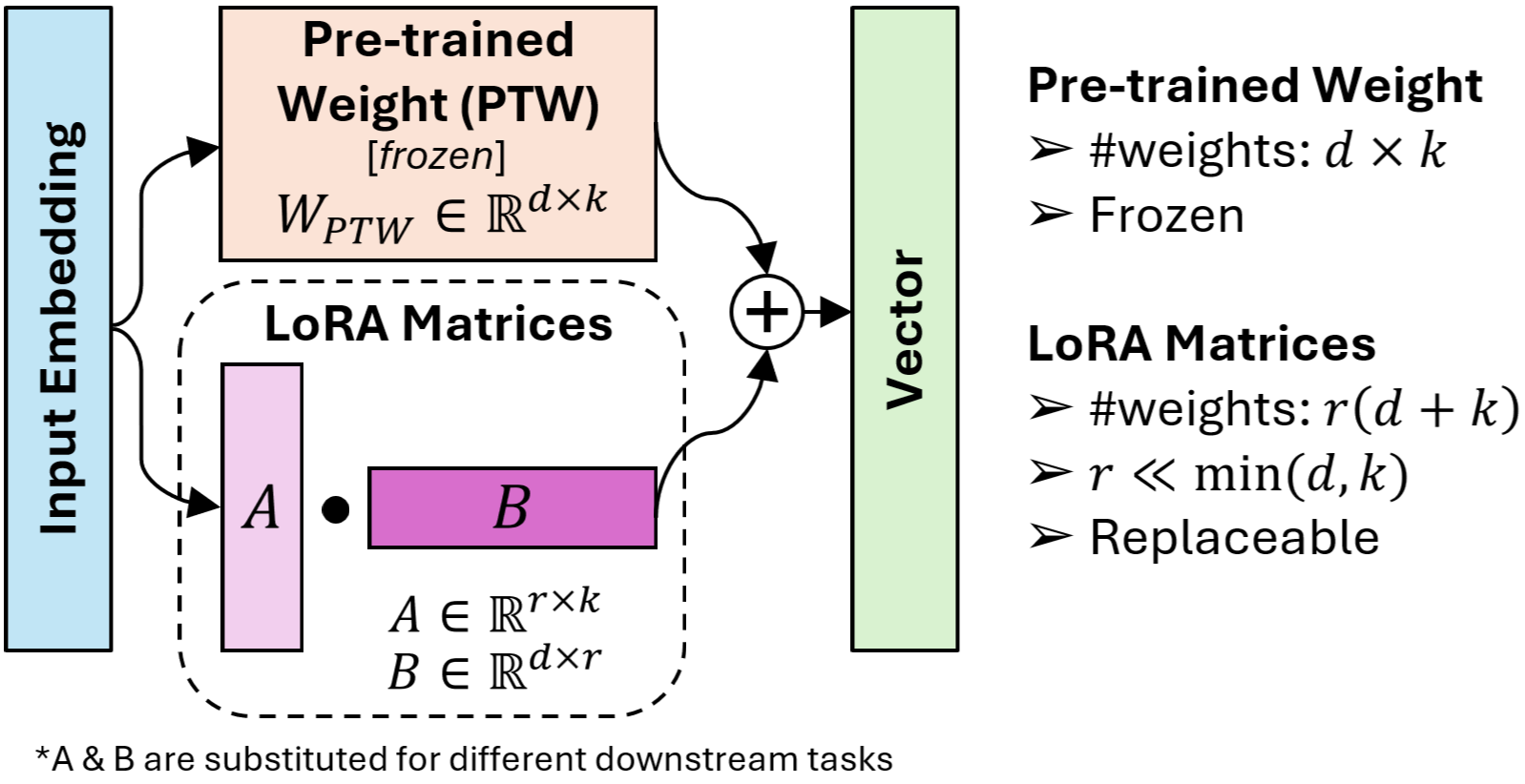}
    \caption{The architecture of \ac{LoRA}.}
    \label{fig:lora_arch}
\end{figure}

\Ac{LoRA} introduces trainable low-rank matrices into specific layers of the transformer architecture without modifying the original model weights as shown in \Cref{fig:lora_arch}.
For specific downstream tasks, only the low-rank matrices are substituted while the original model weights are retained.
\Ac{LoRA} achieves performance comparable to full fine-tuning while significantly reducing computational and storage costs~\cite{lora_origin}.

To exploit the benefits of \ac{LoRA} and improve the energy efficiency and throughput for \ac{LLM} inference, we studied and implemented hardware-software co-design for \ac{LoRA} integrated \ac{LLM} inference accelerator, consisting of heterogeneous compute-in-memory \acp{PE} and \ac{IPCN}.

\section{PRIMAL Hardware Architecture}
\begin{figure*}[t]
    \centering
    \includegraphics[width=1\linewidth]{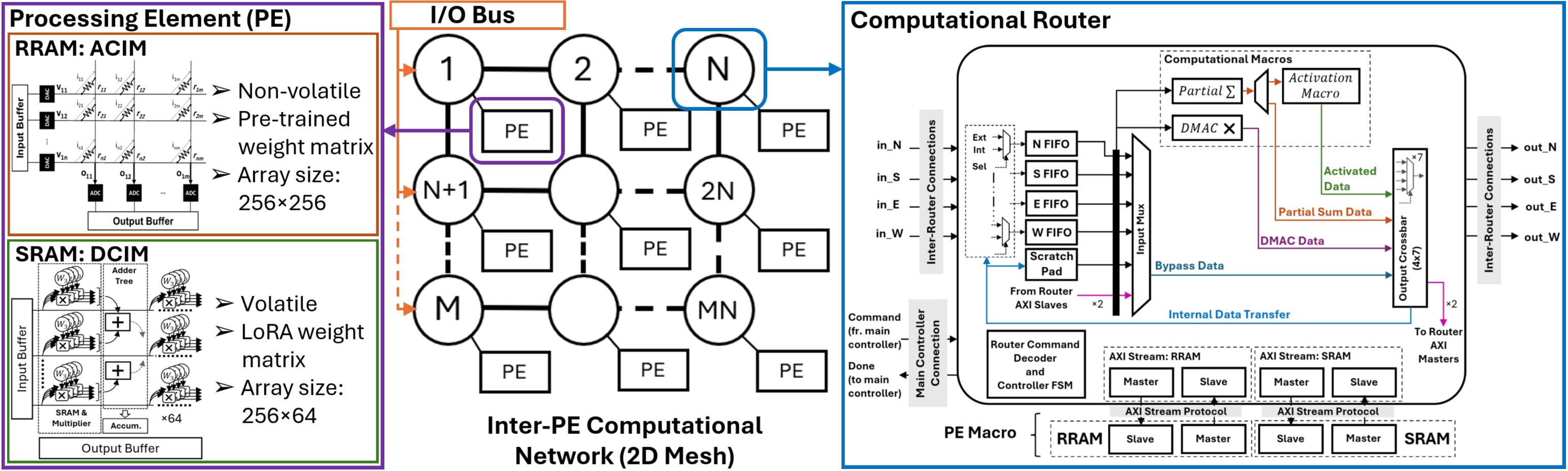}
    \caption{Design of PRIMAL hardware (per compute tile): heterogeneous \ac{PE}, 2D-Mesh \ac{IPCN}, computational router.}
    \label{fig:primal_ct_arch}
\end{figure*}

The PRIMAL hardware architecture adopts a chiplet-based design, where each chiplet is referred to as a \ac{CT}.
Each \ac{CT} integrates a 2D-mesh \ac{IPCN} that interconnects multiple heterogeneous \acp{PE} as shown in \Cref{fig:primal_ct_arch}.
The \ac{IPCN} is responsible for dataflow control and executes \ac{DMAC} operations, representing $\mathbf{Q} \cdot \mathbf{K^T}$ of attention score.
The \acp{PE} perform \ac{SMAC} operations, which correspond to matrix-vector multiplication in the Transformer~\cite{transformer_arch} architecture.

\subsection{Processing Element (PE)}
The \ac{PE} consists of a non-volatile \ac{RRAM-ACIM} macro and a volatile \ac{SRAM-DCIM} macro for \ac{SMAC} of pre-trained weight matrices and \ac{LoRA} matrices respectively.

\subsubsection{RRAM-ACIM}
The \ac{RRAM-ACIM} macro~\cite{rram_acim} exhibits high density and non-volatility, making it particularly suitable for storing large, frozen pre-trained weight matrices.
Its non-volatile nature allows the weights to be programmed only once for a base model, thereby significantly minimizing reconfiguration overhead.
Once initialized, \ac{RRAM-ACIM} performs \ac{SMAC} operations directly in the analog domain, harnessing the inherent parallelism and energy efficiency of analog in-memory computing.

\subsubsection{SRAM-DCIM}
The \ac{SRAM-DCIM} macro~\cite{sram_dcim} offers fast write operations and highly accurate digital MAC operations, making it well-suited for accommodating \ac{LoRA} matrices.
Despite its lower density compared to \ac{RRAM-ACIM}, this limitation is acceptable given the significantly reduced parameter count in \ac{LoRA}.
Moreover, the modular nature of \ac{LoRA}, where matrices are frequently replaced to adapt to different downstream tasks, necessitates rapid reconfiguration.
The fast programmability of SRAM enables efficient updates to \ac{LoRA} weights, thereby minimizing operational stalls and enhancing system throughput and responsiveness.

\subsection{Inter-PE Computational Network (IPCN)}
The 2D-mesh \ac{IPCN} is designed to facilitate seamless integration of compute-in-memory PE macros, enabling efficient orchestration of dataflow across the network to support diverse AI workloads.
Computational tasks such as \ac{DMAC} operations, partial sum accumulation, and activation operations are executed within the routers to enhance processing efficiency.
To accommodate dynamic workload requirements, \ac{IPCN} incorporates a dedicated instruction set architecture that enables re-programmable control over data movement and computation, ensuring adaptability and scalability.

\begin{figure}[t]
    \centering
    \includegraphics[width=0.55\linewidth]{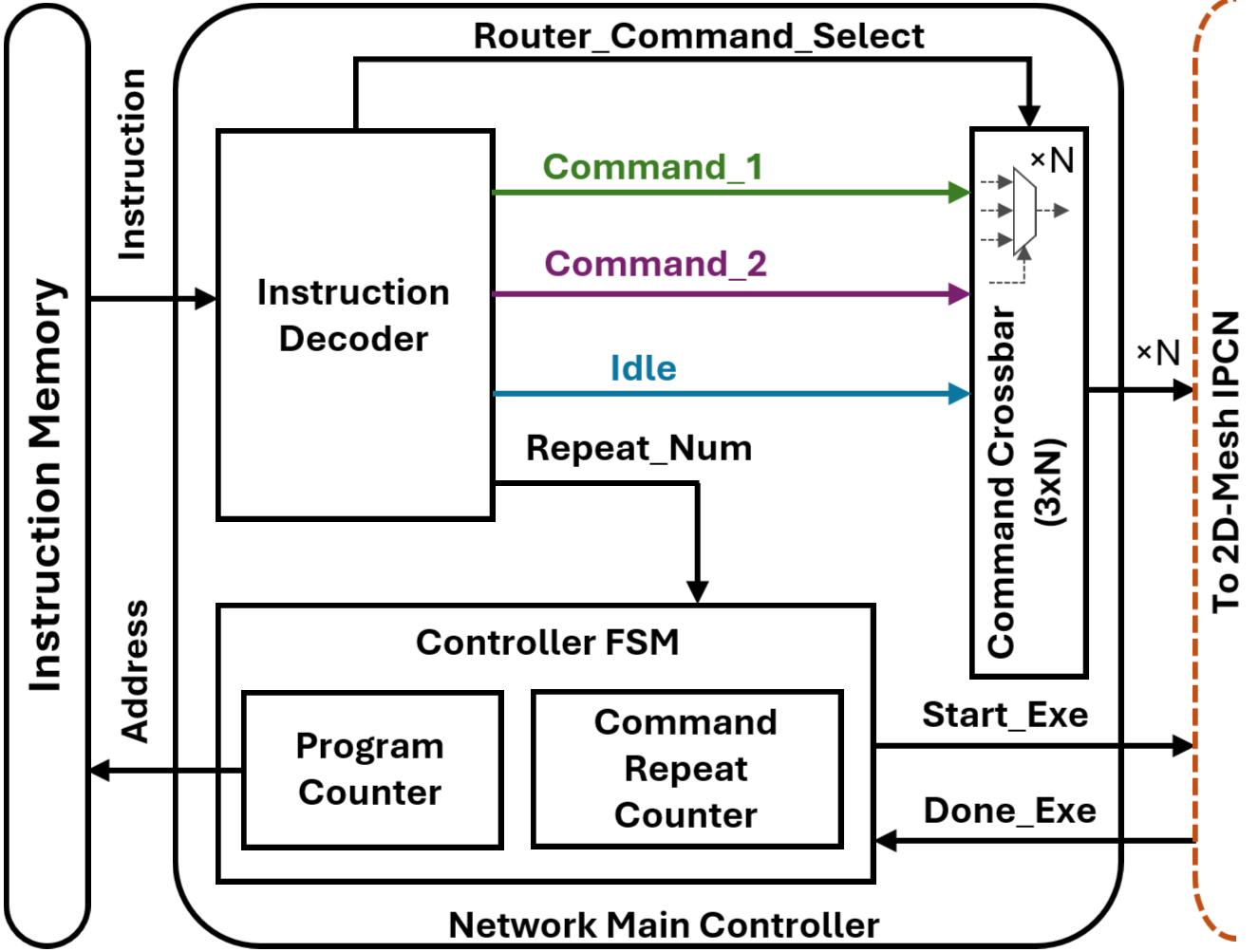}
    \caption{The network main controller (NMC).}
    \label{fig:nmc}
\end{figure}

The unit router features four planar ports dedicated to inter-router communication, along with two pairs of AXI-Stream adapters that interface with \acp{PE} with RRAM and SRAM, respectively.
Each communication port incorporates a FIFO buffer to facilitate temporary data storage and ensure smooth data flow.
To support flexible data handling, the router enables internal data transfers between its buffers and macros.
This capability allows efficient intra-router data movement, optimizing performance for various computational tasks.
An integrated internal controller manages command reception and status reporting, enabling seamless coordination with the \ac{NMC}.

The \ac{NMC}, shown in \Cref{fig:nmc}, governs data movement within the 2D-mesh \ac{IPCN} and coordinates operation of the routers to establish dedicated dataflows optimized for specific \ac{LLM} workloads, via a dedicated instruction set stored in the instruction memory.
Due to operation redundancy in \ac{LLM} workloads, each command to the routers is repeatable as governed by the controller via the instruction.

\section{Mapping and Dataflow Orchestration}

\begin{figure}[t]
    \centering
    \includegraphics[width=0.8\linewidth]{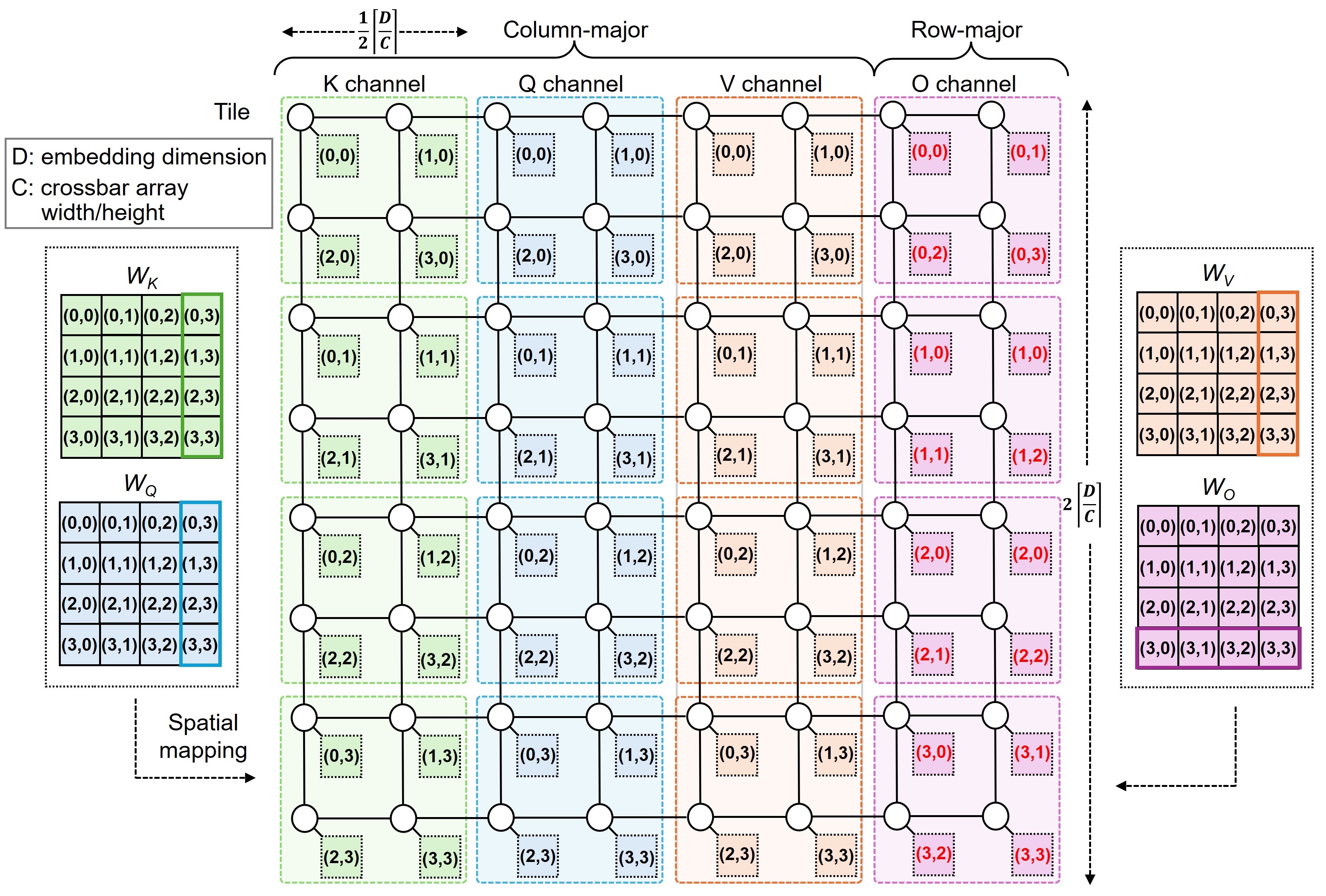}
    \caption{The spatial mapping of weight matrices of an attention layer.}
    \label{fig:mapping}
\end{figure}

\subsection{Mapping}
The partitioned weight matrices $\mathbf{W_Q}$,$\mathbf{W_K}$,$\mathbf{W_V}$, and $\mathbf{W_O}$ are spatially mapped onto the \ac{PE} crossbar arrays, while the corresponding intermediate matrices $\mathbf{Q}$, $\mathbf{K}$, $\mathbf{V}$, and $\mathbf{O}$ (query, key, value, and output, respectively) are allocated to distributed scratchpad memory.
For weight placement on \ac{PE} crossbars, each matrix is heuristically constrained to a column-wise rectangular region (\Cref{fig:mapping}).
The mapping process is optimized by tuning three key factors: intra-matrix shape, inter-matrix shape, and row–column ordering.
Intermediate data are co-located with their associated weights to minimize communication overhead.
Specifically, each intermediate matrix is stored in the scratchpad region corresponding to the \ac{PE} array where its weight matrix resides.
For example, $\mathbf{Q}$ is placed in the scratchpads of the routers attaching to where $\mathbf{W_Q}$ has been pre-mapped.
This co-location strategy enables localized output reduction and improves data reuse efficiency. 

Both pre-trained and \ac{LoRA} weights adopt the same mapping strategy, as \ac{LoRA} introduces low-rank adaptation matrices that are structurally aligned with the original weight matrices and hence, allows the same spatial partitioning and placement strategy to be applied without additional constraints.

\begin{figure}[t]
    \centering
    \includegraphics[width=0.75\linewidth]{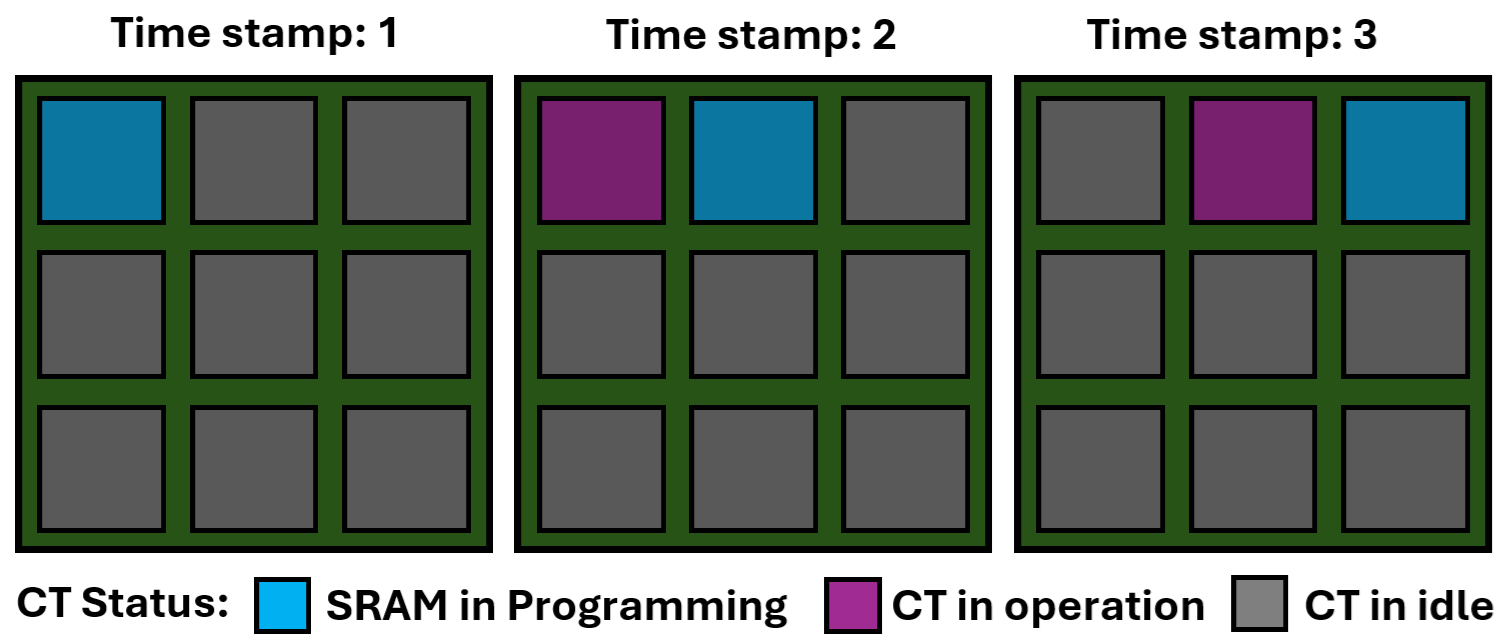}
    \caption{Our Proposed SRAM Reprogramming and Power Gating (SRPG) scheme.}
    \label{fig:srpg}
\end{figure}

\begin{table}[t]
    \caption{System Parameters}
    \centering
    \setlength{\tabcolsep}{7pt}
    \begin{tabular}{c|c|c|c}
         \hline
         \multicolumn{4}{c}{\textbf{System Level}} \\
         \hline
         Bit-width & 64 & Frequency & 1 GHz \\
         \hline \hline
         \multicolumn{4}{c}{\textbf{Compute Tile Level}} \\
         \hline
         IPCN Dimension & 32$\times$32 & PE \# & 1024 \\
         \hline \hline
         \multicolumn{4}{c}{\textbf{Macro Level (per unit Router-PE pair)}} \\
         \hline
         RRAM-ACIM Array & 256$\times$256 & SRAM-DCIM Array & 256$\times$64 \\ 
         \hline
         Scratchpad Size & 32 KB & FIFO Size (each) & 128 B\\ 
         \hline
         DMAC \# & 16 & I/O Pairs \# & 6 \\ 
         \hline
    \end{tabular}    
    \label{tab:sys_param}
\end{table}

\subsection{Dataflow}
The dataflow in \ac{LLM} execution consists of three primary patterns: broadcast, reduction, and unicast.
Initially, input embeddings are broadcast to the \ac{PE} hosting $\mathbf{W_Q}$,$\mathbf{W_K}$, and $\mathbf{W_V}$ for parallel computation.
This enables parallel computation of the $\mathbf{Q}$, $\mathbf{K}$, and $\mathbf{V}$ projections across multiple \acp{PE}.
Since the weights are spatially distributed across multiple columns, a reduction phase aggregates the partial \ac{SMAC} results from multiple \acp{PE}.
Subsequently, unicast operations are employed to compute attention scores by performing $\mathbf{Q} \cdot \mathbf{K^T}$ using \ac{DMAC}, followed by softmax activation.
Unicast ensures that data is delivered point-to-point, reducing unnecessary traffic during the pairwise dot-product computation.

For \ac{KV} cache, during the decode phase, the $\mathbf{K}$ and $\mathbf{V}$ vectors associated with each generated token are appended to statically pre-allocated scratchpad buffers with pre-stored $\mathbf{K/V}$ during prefill.
These buffers are organized in a cyclic fashion across distributed memory units, enabling uniform load distribution and mitigating memory contention.
The cyclic placement strategy ensures that scratchpad utilization remains balanced irrespective of sequence length, thereby sustaining throughput and minimizing latency in long-context inference scenarios.

The collective communication pattern is orchestrated using a spanning tree algorithm, which determines the routing paths for each phase.
This algorithm ensures balanced and congestion-free traffic by leveraging the regular and aligned mapping of weights and intermediate data.

\begin{table}[t]
    \caption{PRIMAL Benchmarking: Throughput and Power}
    \centering
    \setlength{\tabcolsep}{2.5pt}
    \begin{tabular}{c|c|c|c|c|c}
         \hline
         \multirow{2}{*}{Model}
         & LoRA* & Context Length & Throughput & Average & Efficiency \\
         & Matrices & (Input/Output) & (tokens/s) & Power (W) & (tokens/J) \\ 
         \hline
         \multirow{4}{*}{\shortstack{Llama 3.2 \\ 1B}}
         & \multirow{2}{*}{Q} & 1024/1024 & 966.32 & \multirow{4}{*}{2.23} & 433.33 \\
         & & 2048/2048 & 565.46 & & 253.57 \\
         \cline{2-4} \cline{6-6}
         & \multirow{2}{*}{Q, V} & 1024/1024 & 963.47 & & 432.04 \\
         & & 2048/2048 & 564.48 & & 253.13 \\
         \hline
         \multirow{4}{*}{\shortstack{Llama 3 \\ 8B}}
         & \multirow{2}{*}{Q} & 1024/1024 & 308.76 & \multirow{4}{*}{9.58} & 32.23 \\
         & & 2048/2048 & 221.37 & & 23.11 \\
         \cline{2-4} \cline{6-6}
         & \multirow{2}{*}{Q, V} & 1024/1024 & 307.89 & & 32.12 \\
         & & 2048/2048 & 220.77 & & 23.04 \\
         \hline
         \multirow{4}{*}{\shortstack{Llama 2 \\ 13B}}
         & \multirow{2}{*}{Q} & 1024/1024 & 191.68 & \multirow{4}{*}{14.76} & 12.99 \\
         & & 2048/2048 & 145.81 & & 9.88 \\
         \cline{2-4} \cline{6-6}
         & \multirow{2}{*}{Q, V} & 1024/1024 & 190.98 & & 12.94 \\
         & & 2048/2048 & 145.40 & & 9.85 \\
         \hline
    \end{tabular}
    
    \vspace{0.15cm}
    {\raggedright *LoRA with Rank 8 \\}
    \label{tab:benchmark_tp_pwr}
\end{table}

\begin{table}[t]
    \caption{PRIMAL Latency: TTFT and ITL}
    \centering
    \setlength{\tabcolsep}{10pt}
    \begin{tabular}{c|c|c|c|c}
         \hline
         \multirow{2}{*}{Model}
         & LoRA* & Context Length & TTFT & ITL \\
         & Matrices & (Input/Output) & ($s$) & ($ms$) \\ 
         \hline
         \multirow{4}{*}{\shortstack{Llama 3.2 \\ 1B}}
         & \multirow{2}{*}{Q} & 1024/1024 & 0.370 & 1.708 \\
         & & 2048/2048 & 1.192 & 2.955 \\
          \cline{2-5}
         & \multirow{2}{*}{Q, V} & 1024/1024 & 0.373 & 1.711 \\
         & & 2048/2048 & 1.199 & 2.958 \\
         \hline
          \multirow{4}{*}{\shortstack{Llama 3 \\ 8B}}
         & \multirow{2}{*}{Q} & 1024/1024 & 0.710 & 5.726 \\
         & & 2048/2048 & 2.012 & 8.052 \\
          \cline{2-5}
         & \multirow{2}{*}{Q, V} & 1024/1024 & 0.782 & 5.738 \\
         & & 2048/2048 & 2.037 & 8.065 \\
         \hline
          \multirow{4}{*}{\shortstack{Llama 2 \\ 13B}}
         & \multirow{2}{*}{Q} & 1024/1024 & 0.962 & 9.494 \\
         & & 2048/2048 & 2.494 & 12.499 \\
          \cline{2-5}
         & \multirow{2}{*}{Q, V} & 1024/1024 & 0.982 & 9.513 \\
         & & 2048/2048 & 2.533 & 12.518 \\
         \hline
    \end{tabular}
    
    \vspace{0.15cm}
    {\raggedright *LoRA with Rank 8 \\}
    \label{tab:benchmark_ttft_itl}
\end{table}

\subsection{SRAM Reprogramming and Power Gating (SRPG)}
For different downstream tasks, the LoRA matrices must be updated, which necessitates reprogramming the \ac{SRAM-DCIM} macros.
Additionally, \ac{LLM} inference workloads are executed in a strictly sequential, layer-by-layer manner~\cite{iccad_leap}.
During the computation of a given layer, all other layers remain idle, presenting a significant opportunity for power optimization.
To leverage both the sequential execution pattern and the SRAM reprogramming requirements, PRIMAL adopts a \ac{CT}-based, layer-wise weight allocation strategy, which maps each layer to adjacent \acp{CT}.

As \Cref{fig:srpg} shows, at the initialization phase for a distinct downstream task (Time Stamp 1), the SRAMs in the first \ac{CT} are reprogrammed.
Once the first \ac{CT} begins executing its assigned workload, the SRAMs in the next \ac{CT} are reprogrammed in parallel.
For \ac{CT} in idle state, its \ac{IPCN} and \ac{RRAM-ACIM} macros are power-gated to save energy.
However, the SRAMs and scratchpad memory macros remain powered on across all \ac{CT}s to preserve volatile \ac{LoRA} weights and ensure retention of context window data for \ac{KV} caching, respectively.
This operation scheme effectively reduces system power consumption by minimizing energy usage in idle \acp{CT}.

\section{System Evaluation}

The system was evaluated through a comprehensive hardware-software co-verification methodology.
Digital hardware components were designed and verified using Verilog HDL.
Hardware synthesis was performed using \textit{Synopsys Design Compiler}, followed by place-and-route using \textit{Cadence Innovus}.
Power and area of the scratchpad memory macro were obtained using \textit{CACTI}~\cite{cacti}, while other hardware macros were modeled and emulated in software using mathematical abstractions.
Inference emulation and benchmarking were conducted using a cycle-accurate, instruction-level simulator based on the \ac{IPCN} instruction set with the mapping scheme.

\begin{figure}[t]
    \centering
    \includegraphics[width=1\linewidth]{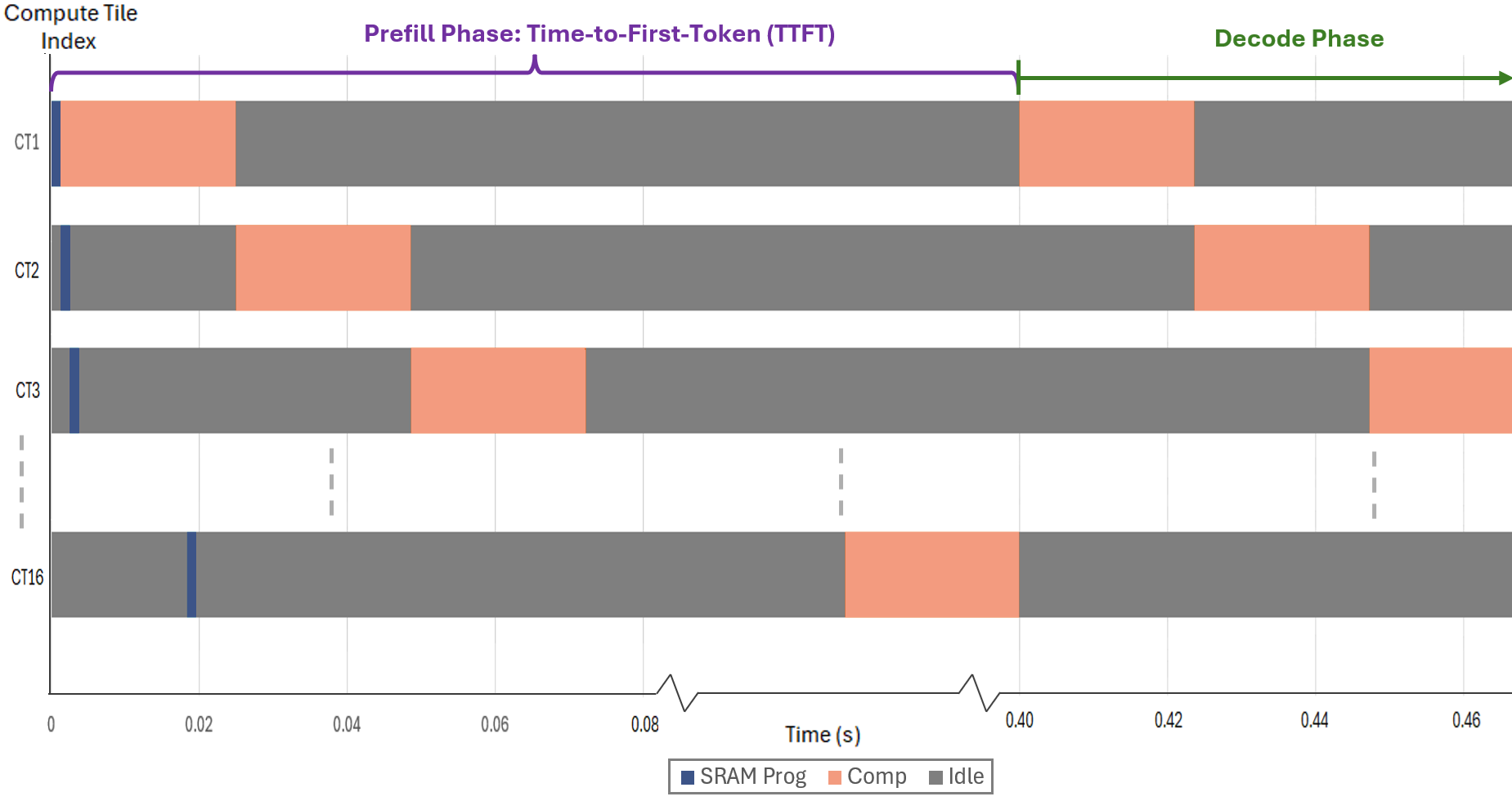}
    \caption{The timing diagram of hardware scheduling for Llama 3.2-1B on PRIMAL.}
    \label{fig:timing_diag}
\end{figure}

\subsection{Performance Benchmarking}
\subsubsection{Throughput and Power}
In PRIMAL, the average system power is dependent on the number of \acp{CT}.
As the model size increases, the power consumption rises due to the activation of additional \acp{CT} required to store and process the expanded model weights.
Concurrently, system throughput degrades as the enlarged model induces greater data movement and computational demands within the \ac{IPCN}. 

Compared to Nvidia H100, PRIMAL achieves $1.5\times$ throughput and $25\times$ energy efficiency (9.85~tks/J vs. 0.4~tks/J) for \ac{LoRA} rank~8 (Q, V) on Llama-13B (2048/2048, Batch~1).

\subsubsection{Time-To-First-Token (TTFT)}
TTFT denotes the latency between the reception of a prompt and the generation of the first output token by the \ac{LLM}.
In PRIMAL, this corresponds to the prefill phase, which includes both the reprogramming of \acp{SRAM-DCIM} in the initial \ac{CT} and the aggregate computation time across all \ac{LLM} layers (\Cref{fig:timing_diag}).
Due to pipelined execution scheme of PRIMAL, \ac{SRPG}, the reprogramming of \acp{SRAM-DCIM} in subsequent \acp{CT} is effectively overlapped with computation and thus, does not contribute to TTFT.
Consequently, TTFT is primarily bounded by the compute time of each \ac{CT}, which serves as the critical path in the pipeline.

\subsubsection{Inter-token Latency (ITL)}
ITL, which corresponds to the decode phase, is primarily determined by the total number of layers in the \ac{LLM}.
As the model size scales, the number of layers increases proportionally, leading to deeper computational pipelines.
Consequently, data must propagate through more layers that are mapped across additional \acp{CT} in PRIMAL, thereby incurring higher latency between successive token generations (\Cref{tab:benchmark_ttft_itl}).

\subsection{Hardware Scalability}
The implementation of \ac{SRPG} scheme enables power gating of \ac{IPCN} and \ac{RRAM-ACIM} macros within \acp{CT} in idle states, thereby substantially reducing overall system power consumption.
Empirical evaluation on the various-size models demonstrates that \ac{SRPG} achieves up to 80\% power savings compared to the baseline configuration without power gating.
Consequently, \ac{SRPG} ensures that system power scales sub-linearly with respect to the \ac{LLM} size (\Cref{tab:benchmark_tp_pwr}), making PRIMAL highly scalable for running larger \acp{LLM}.

\subsection{Macro Power and Area}
The average power and area breakdown of the PRIMAL macros are summarized in \Cref{tab:pwr_and_area}.
The \ac{RRAM-ACIM} macro dominates area usage due to integrated analog \ac{SMAC} operations and the need for DACs and ADCs.
In contrast, the \ac{SRAM-DCIM} macro, which performs digital \ac{SMAC} using adder trees and shifters, eliminates converters and resulting in lower area but higher power consumption due to increased digital switching activity and less efficient digital computation.

\begin{table}[t]
    \caption{Avg. Power \& Area Breakdown of Hardware Macros (Unit)}
    \centering
    \setlength{\tabcolsep}{3.3pt}
    \begin{tabular}{c|cc|cc}
         \hline
         Macro & Power ($u$W) & Breakdown & Area (mm$^2$) & Breakdown \\
         \hline
         RRAM-ACIM \cite{rram_acim} & 120 & 9.9\% & 0.1442 & 65.2\% \\
         \hline
         SRAM-DCIM \cite{sram_dcim} & 950 & 78.1\% & 0.035 & 15.8\% \\
         \hline
         Scratchpad Mem. & 42 & 3.5\% & 0.013 & 5.9\% \\
         \hline
         Router & 103 & 8.5\% & 0.029 & 13.1\% \\
         \hline \hline
         Total & \multirow{2}{*}{1215} & \multirow{2}{*}{100\%} & \multirow{2}{*}{0.2212} & \multirow{2}{*}{100\%} \\
         (Router-PE pair) & & & & \\
         \hline
    \end{tabular}

    \vspace{0.1cm}
    {\raggedright $^{\#}$Technology node: 7~nm; Area per \ac{CT} Chiplet: 227.5 mm$^2$ \\}
    \label{tab:pwr_and_area}
\end{table}

\section{Conclusion}
Integrated with \ac{IPCN} interconnect and adaptability of \ac{LoRA}, PRIMAL leverages the energy efficiency of CIM to deliver energy efficient \ac{LLM} inference while enabling seamless downstream task adaptation.
Furthermore, the proposed \ac{SRPG} mechanism ensures efficient SRAM reprogramming and optimizes the compute pipeline, achieving sub-linear power scaling with respect to \ac{LLM} size.
Evaluation shows it achieves $1.5\times$ throughput and $25\times$ energy efficiency improvement
over Nvidia H100 for Llama-13B (2048/2048, LoRA rank~8 Q,V).
Our design paradigm not only addresses the escalating energy demands of large-scale \acp{LLM} but also establishes PRIMAL as a highly scalable architecture, capable of supporting larger \acp{LLM} with \ac{LoRA} at minimal power overhead via \ac{SRPG}.

\bibliographystyle{IEEEtran}
\bibliography{references}

\end{document}